\documentclass[aps,prd,showpacs,preprintnumbers]{revtex4-2}
\usepackage{amsmath}
\usepackage{subfig}
\usepackage{graphicx}
\usepackage{epstopdf}
\usepackage{amssymb,amsfonts}
\usepackage{latexsym}
\usepackage{epsfig}
\usepackage{amssymb}
\usepackage{color}
 \usepackage{bm}
\newcommand{\be}{\begin{eqnarray}}
\newcommand{\ee}{\end{eqnarray}}

\newcommand{\bea}{\begin{eqnarray}}
\newcommand{\eea}{\end{eqnarray}}

\begin{document}
\title{Effect of the anomalous magnetic moment of quarks on magnetized QCD matter and meson spectra}
\author{Kun Xu$^{1,2}$ }
\thanks{xukun@mail.ihep.ac.cn}
\author{Jingyi Chao$^{3}$}
\author{Mei Huang$^{1}$}
\thanks{huangmei@ucas.ac.cn}
\affiliation{$^{1}$ School of Nuclear Science and Technology, University of Chinese Academy of Sciences, Beijing 100049, China}
\affiliation{$^{2}$ Institute of High Energy Physics, Chinese Academy of Sciences, Beijing 100049, P.R. China}
\affiliation{$^{3}$ Institute of Modern Physics, Chinese Academy of Sciences,  Lanzhou 730000, P.R. China }

\begin{abstract}
We systematically investigate the effect of the anomalous magnetic moment(AMM) of quarks on the magnetized QCD matter, including the magnetic susceptibility, the inverse magnetic catalysis around the critical temperature and the neutral/charged pion and rho meson spectra under magnetic fields. The dynamical AMM of quarks, its coupling with magnetic field causes Zeeman splitting of the dispersion relation of quarks thus changes the magnetism properties and meson mass spectra under magnetic fields. It is found that including the AMM of quarks cannot fully understand lattice results of the magnetized matter: The AMM of quarks reduces the dynamical quark mass thus causes the inverse magnetic catalysis around $T_c$.  The neutral pion mass is very sensitive to the AMM, it decreases with magnetic field quickly, and the charged pion mass shows a nonlinear behavior, i.e., firstly linearly increases with the magnetic field and then saturates at strong magnetic field.  For rho meson, it is observed that AMM reduces the mass of neutral rho meson mass with different $s_z$, and reduces the mass of $s_z=+1,0$ component charged rho meson mass but enhances the $s_z=-1$ component charged rho meson mass.  The magnetic susceptibility at low temperature can be either positive or negative with different AMM.  
\end{abstract}
\maketitle

\section{Introduction}
Understanding properties of QCD matter under strong magnetic filed is of vital importance to further explore the interior of magnetar~\cite{Duncan:1992hi,Tatsumi:2006wr}, neutron-star merges~\cite{Kiuchi:2015sga,Baiotti:2016qnr}, non-central heavy-ion collisions~\cite{Skokov:2009qp,Deng:2012pc} and the evolution of the early universe~\cite{Vachaspati:1991nm}. The study of the  QCD vacuum and strongly interacting matter under external magnetic fields has attracted much attention, see reviews, e.g. Refs.~\cite{Andersen:2014xxa,Miransky:2015ava,Huang:2015oca,Kharzeev:2015znc,Bzdak:2019pkr}.  With the presence of a magnetic field background, the strongly interacting matter shows a large number of exotic phenomena, for example, Chiral Magnetic Effect(CME)~\cite{Kharzeev:2007tn,Kharzeev:2007jp,Fukushima:2008xe,KharzeevSon:2010gr}, Magnetic Catalysis(MC) in the vacuum~\cite{Klevansky:1989vi,Klimenko:1990rh,Gusynin:1995nb}, Inverse Magnetic Catalysis(IMC) around the critical temperature~\cite{Bali:2011qj,Bali:2012zg,Bali:2013esa}.

The catalysis of the chiral symmetry breaking induced by the magnetic field, i.e., the MC effect can be easily understood from the dimension reduction. On the other hand, the IMC effect, the critical temperature of chiral phase transition decreases with the magnetic field, which intuitively contradicts with the MC effect and still remains as a puzzle, though there have been many literatures trying to explain the IMC by considering neutral pion fluctuations~\cite{Fukushima:2012kc}, chirality imbalance~\cite{Chao:2013qpa}, running coupling constant~\cite{Ferrer:2014qka}. Recently, lattice calculations show more interesting and novel properties of magnetized QCD matter: The charged pion mass shows a non-monotonic magnetic field dependent behavior~\cite{Ding:2020jui}, and magnetized matter exhibits diamagnetism (negative susceptibility) at low temperature and paramagnetism (positive susceptibility) at high temperature~\cite{Bali:2012jv,Bali:2020bcn}.

Magnetic fields modify the spectrum of charged particles. The point particle approximation gives the charged pion mass $m^2_{\pi^{\pm}}(B)=m^2_{\pi^{\pm}}(B=0)+eB$ increasing linearly  with the magnetic field, and neutral pion mass keeps as a constant. Calculation in the effective quark model, e.g, the Namu--Jona-Lasinio (NJL) model which takes into account of quark magnetization, modifies the linear slope of charged pion and shows similar results for pion mass spectra as point-approximation results ~\cite{Liu:2014uwa,Liu:2015pna,Liu:2018zag,Wang:2017vtn,Mao:2018dqe,Avancini:2015ady,Avancini:2016fgq,Coppola:2018vkw,Coppola:2019uyr,Fayazbakhsh:2012vr,Fayazbakhsh:2013cha}. In the NJL model, mesons are considered as quantum fluctuations in Random Phase Approximation(RPA), where the mesons are introduced via the summation of an infinite number of quark loops~\cite{Wang:2017vtn,Mao:2018dqe,Coppola:2018vkw,Coppola:2019uyr,Klevansky:1992qe,He:1997gn,Rehberg:1995nr}. However, with presence of magnetic field, the Schwinger phase appears in each quark propagator~\cite{Schwinger:1951nm}. For neutral pion, the Schwinger phases cancel out for each loop, while they cannot for charged pion. 
In Ref.\cite{Coppola:2018vkw,Coppola:2019uyr} the authors employed the Ritus eigenfunction method in the two-flavor NJL model, which allows to properly take into account the presence of Schwinger phases in the quark propagators. They found that in the region $eB\sim 0-1.5\text{GeV}^2$, neutral pion mass decreases slightly while charged pion mass steadily increases. And in Ref.\cite{Wang:2017vtn}, it is found that the charged pion becomes much heavier in the magnetic field and are sensitive to the field strength, while the neutral pion still keeps as a Nambu-Goldstone particle in the region around $eB \sim 0-0.4 \text{GeV}^2$. And Ref.\cite{Avancini:2015ady} shows that neutral pion mass firstly decreases and then increases with magnetic field, which is consistent with lattice result in Ref.\cite{Hidaka:2012mz,Andreichikov:2013zba}. Similar results are also found in another lattice 
calculation Ref.\cite{Bali:2017ian}. In Ref.\cite{Bali:2011qj}, charged pion mass was calculated in Lattice QCD, and they found that charged pion mass increases with magnetic field in the region of $eB \sim 0 - 0.3\text{GeV}^2$, however, for $eB>0.3\text{GeV}^2$, charged pion mass doesn't show a increase trend with magnetic field. Recent lattice calculation in Ref.\cite{Ding:2020jui} shows that the neutral pion mass decreases with the magnetic field while the charged pion and kaon show non-monotonic behaviors, the mass of which firstly increases linearly and then decreases as magnetic field increases, and all these masses show a saturation at $eB \gtrsim  2.5 \text{GeV}^2$, which are quite different from point-particle approximation and previous results from effective models. It is worthy of mentioning that lattice results in \cite{Bali:2011qj,Bali:2017ian} show a nonlinear and saturation behavior for charged pion, no decreasing with magnetic field is observed at strong magnetic field.

The remaining IMC puzzle and recently discovered properties of magnetized matter attract our renewed interest to revisit QCD vacuum and matter under external magnetic field and try to find the underline mechanism for these properties. It is known that the dynamical chiral symmetry broken is one of the most significant feature of QCD, where quarks obtain a dynamical mass. It has been found that anomalous magnetic moments(AMM) of quarks can also be generated dynamically along with dynamical quark mass~\cite{Chang:2010hb,Ferrer:2008dy,Ferrer:2013noa}. In Ref.\cite{Chang:2010hb} the authors explained how dynamical chiral symmetry breaking produces a dressed light-quark with a momentum-dependent anomalous chromomagnetic moment as well as the generation of an anomalous electromagnetic moment for the dressed light-quark in QED and QCD. In Ref.\cite{Ferrer:2013noa}, the authors considered one-flavor NJL model with a new channel $\sim (\bar{\psi}\Sigma^{3}\psi)^2 + (\bar{\psi}i\gamma^5 \Sigma^{3}\psi)^2$ in the presence of magnetic field, and they found that the chiral condensate and new condensate $ \langle \bar{\psi}i\gamma^1\gamma^2\psi \rangle $, which corresponds to AMM for quarks, are different from zero at vacuum with magnetic field simultaneously. A two-flavor NJL model with tensor channel was also investigated in Ref.\cite{Mao:2018jdo}, and similar results were obtained. And in Ref.\cite{Ferrer:2008dy,Ferrer:2009nq}, the authors investigated the non-perturbative generation of an anomalous magnetic moment for massless fermions in the presence of an external magnetic field, and they proved that the phenomenon of magnetic catalysis of chiral symmetry breaking is also responsible for the generation of the dynamical anomalous magnetic moment. Thus, once the quarks acquire a dynamical mass, they should also acquire a dynamical AMM~\cite{Ferrer:2013noa,Bicudo:1998qb,Chang:2010hb,Ferrer:2009nq,Ferrer:2008dy}. 

There have been some papers working on how AMM of quarks influences QCD phase diagram as well as mesonic properties~\cite{Chaudhuri:2019lbw,Ghosh:2020xwp,Strickland:2012vu,Chaudhuri:2020lga,Ferrer:2015wca,Fayazbakhsh:2014mca,Ferrer:2014qka,Chaudhuri:2020lga,Ghosh:2020xwp,Chao:2020jjy}. For example, in Ref.\cite{Chaudhuri:2019lbw}, the authors used two-flavor NJL model with AMM and found that the critical temperature for chiral transition decreases with the external magnetic field, while a sudden jump for pion mass at and above the Mott transition temperature appears when the AMM of the quarks are taken into consideration. And Ref.\cite{Fayazbakhsh:2014mca} found that Inverse Magnetic Catalysis occurs for large enough AMM. In Ref.\cite{Ferrer:2014qka} the authors considered NJL model with dynamical induced AMM and found that with magnetic-dependent coupling constants, the inverse magnetic catalysis can be obtained.

In the present paper, we focus on the dynamical quark mass as well as meson mass, e.g., pion and rho, in the presence of magnetic field with anomalous magnetic moments of quarks. This paper is organized as follows: in Sec. \ref{sec-model} we introduce two-flavor NJL model with AMM in the presence of magnetic field, and investigate the effect of AMM on dynamical quark mass and the magnetism property. Then we investigate pion and rho mass as a function of magnetic field with different AMM In Sec.\ref{sec-meson}. Finally, we discuss the results in Sec.\ref{sec-con}. 

\section{MODEL SETUP}
\label{sec-model}
We choose two-flavor Nambu--Jona-Lasinio model including the AMM of quarks in the presence of magnetic field, the Lagrangian of which takes form of~\cite{Chaudhuri:2019lbw,Chaudhuri:2020lga,Fayazbakhsh:2014mca}:
\bea
	\label{equ:L}
	\begin{split}
	\mathcal{L}  =\bar{\psi}(i\gamma^{\mu}D_{\mu}-m_0+\kappa_f q_fF_{\mu\nu}\sigma^{\mu\nu})\psi & + G_S{\Big \{}(\bar{\psi}\psi)^2+(\bar{\psi}i\gamma^{5}\vec{\tau}\psi)^2 {\Big \}}\\
	& - G_V{\Big \{}(\bar{\psi}\gamma^{\mu}\vec{\tau}\psi)^2+(\bar{\psi}\gamma^{\mu}\gamma^{5}\vec{\tau}\psi)^2{\Big \}}.
	\end{split}
\eea
Here $\psi$ are two-flavor quark filed $\psi=(u,d)^{T}$,  $m_0$ is current mass and we assume the current quark mass for both flavors are the same: $m_u=m_d=m_0$. The covariant derivative $D_{\mu}=\partial_{\mu}-i q_f A_{\mu}$ with $q_f$ the electric charge of quarks, and $A_{\mu}$ is the Abelian gauge field and field strength $F_{\mu\nu}=\partial_{\mu}A_{\nu}-\partial_{\nu}A_{\mu}$. Without loss of generality, we choose external uniform magnetic field along z-direction, which led to $A_{\mu}=\{0,0,Bx,0\}$. The term $\bar{\psi}\kappa_f q_fF_{\mu\nu}\sigma^{\mu\nu}\psi$, with $\sigma^{\mu\nu}=\frac{i}{2}[\gamma^{\mu},\gamma^{\nu}]$, is to present the contribution of AMM, and $\kappa_f$ is defined as $\kappa_f=\alpha_f \mu_B$ with $\mu_B=\frac{e}{2M}$ the Bohr magneton, and $M$ the constituent quark mass as defined below. At one-loop level we have $\alpha_f=\frac{\alpha_e q_f^2}{2\pi}$, with $\alpha_e=\frac{1}{137}$ the electromagnetic fine structure constant. However, to study how the AMM influence quark mass as well as meson mass, we treat $\kappa_f$ as a free parameter and be flavor-independent, e.g.,$\kappa_u=\kappa_d=\kappa$. Besides, $G_S$ and $G_V$ are the coupling constants for (pseudo-)scalar and (pseudo-)vector interaction channel, respectively. Then the Lagrangian after mean-field approximation is given by:
\begin{equation}
\mathcal{L}=-\frac{(M-m_0)^2+\pi^a \pi^a}{4G_S}+\frac{V_{\mu}^aV^{\mu,a}+A_{\mu}^aA^{\mu,a}}{4G_V}+\bar{\psi}(i\gamma^{\mu}D_{\mu}-M+\kappa q_fB\sigma^{12})\psi,
\end{equation}
where we define:
\be
	\begin{split}
		M=m_0-2G_S \langle\bar{\psi}\psi\rangle, & \quad \vec{\pi}=-2G_S \langle\bar{\psi}i\gamma^5 \vec{\tau}\psi\rangle , \\
		V_{\mu}^a=-2G_V \langle\bar{\psi}\gamma_{\mu}\tau^a\psi\rangle, & \quad A_{\mu}^a=-2G_V \langle\bar{\psi}\gamma_{\mu}\gamma^5\tau^a\psi\rangle .
	\end{split}
\ee
It is know that NJL model is non-renormalized, thus regularization scheme is necessary for finite numerical results, and in this paper, a soft cut-off is applied for momentum integration and Landau level summation during the numerical calculation:
\be
\frac{|q_fB|}{2\pi}\sum_{n}\int \frac{dp_z}{2\pi} \rightarrow \frac{|q_fB|}{2\pi}\sum_{n}\int \frac{dp_z}{2\pi}f_\Lambda (p_z,n)
\ee
with
\be
\label{equ:cutofffun}
f_\Lambda (p_z,n)=\frac{\Lambda^{10}}{\Lambda^{10}+(p_z^2+2n|q_fB|)^{5}}.
\ee  
There are four parameters in total: current quark mass $m_0$, three-momentum cutoff parameter $\Lambda$ in Eq.(\ref{equ:cutofffun}), (pseudo-)scalar/(pseudo-)vector coupling constant $G_S$/$G_V$, and these parameters are determined by fitting to experimental datas at zero temperature and vanishing magnetic field. In this work, we have used two sets of parameters: I), $m_0=5\text{MeV}$, $\Lambda=624.18\text{MeV}$ and $G_S \Lambda^2=2.014$, which corresponds to pion decay constant $f_{\pi}=93\text{MeV}$, pion mass $m_{\pi}=135.6\text{MeV}$ as well as the quark condensate $\langle\bar{\psi} \psi \rangle=-(251.8\text{MeV})^3$, and this set of parameters is used to investigate dynamical quark mass as well as pion mass; II), $m_0=5\text{MeV}$, $\Lambda=582\text{MeV}$, $G_S \Lambda^2=2.388$ and $G_V \Lambda^2=1.73$, which are chosen in such a way that $m_\pi=140\text{MeV}$, $m_{\rho}=768\text{MeV}$ while $M=458\text{MeV}$ at zero temperature as used in Ref.\cite{He:1997gn,Liu:2014uwa}, which indicates the quark condensate $\langle\bar{\psi} \psi \rangle = -(267\text{MeV})^3$. In particular, the dynamical generated quark mass $M$ resulting from parameter set II) is deliberately chosen to be large to avoid the decay process $\rho \rightarrow q\bar{q}$ at zero temperature, and this set is used for the study of rho meson.

\subsection{Dispersion Relation for Fermions with AMM}
\label{app_DIS}
\begin{figure}[h!]
	\centering
	\includegraphics[width=10cm]{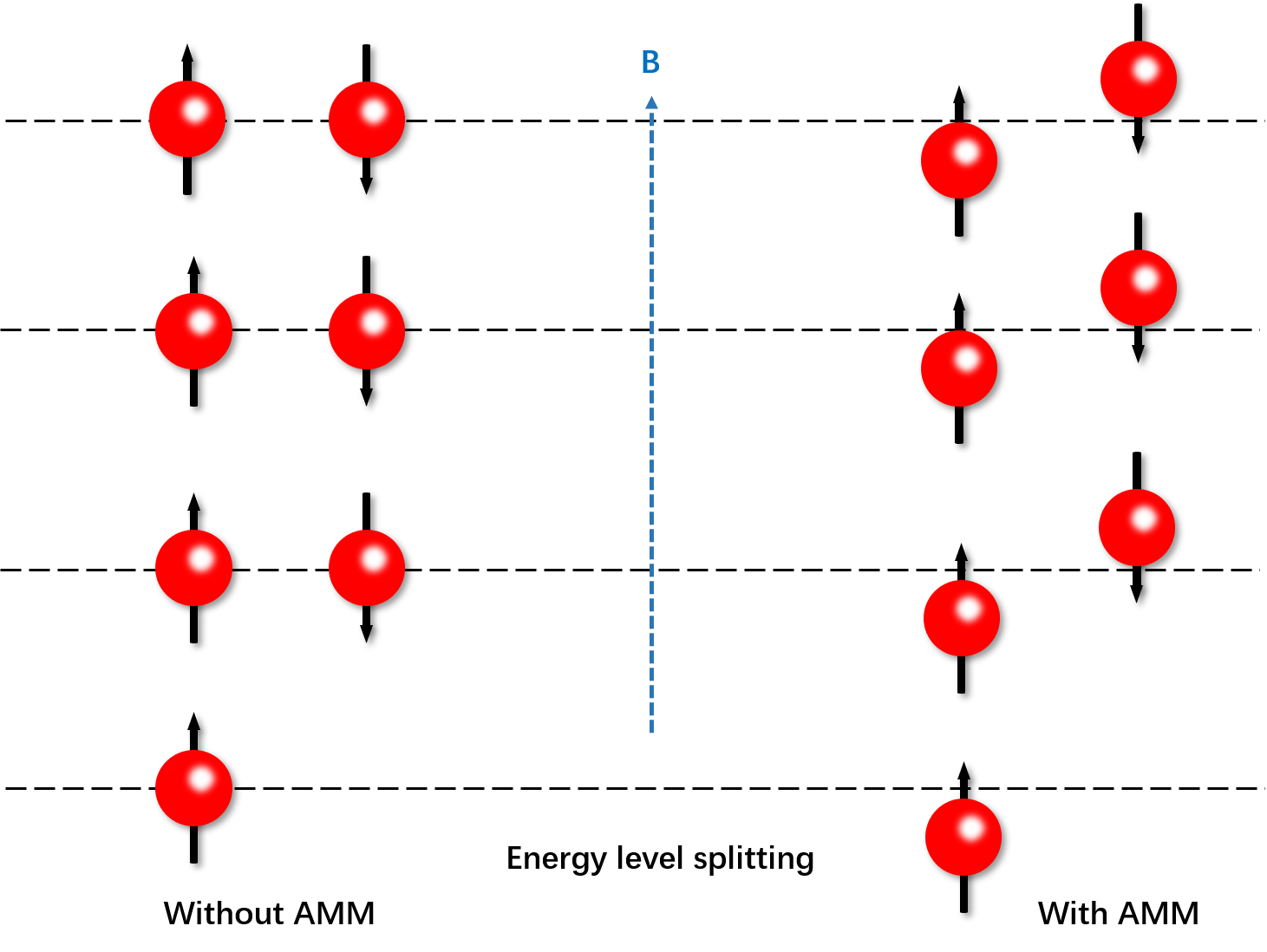}
	\caption{Energy levels of  positive charged fermion for the case with(right) and without(left) AMM in the presence of magnetic field.}
	\label{fig:AMM-dis}
\end{figure}

In this part, we derive the dispersion relation for positive charged fermion $\psi$ with charge $q$, dynamical mass $M$ and the anomalous magnetic moment $\kappa$ in the presence of homogeneous magnetic field $B$, the Dirac equation of which is given by
\be
(i\gamma^{\mu}D_{\mu}-M+\frac{1}{2}\kappa q \sigma^{\mu\nu}F_{\mu\nu})\psi=0.
\ee
Similar with Eq.(\ref{equ:L}), we take the magnetic field along $z$ direction, and to simplify following derivation, we set $T=\kappa q B$, then the Dirac equation becomes:
\be\label{equ_diraequ}
(i\gamma^{\mu}D_{\mu}-M+ T\sigma^{12})\psi=0,
\ee
where $\sigma^{12}=i\gamma^1 \gamma^2$. In this case, the general solution of $\psi$ has form of:
\be\label{equ_genesolu}
\psi=\text{e}^{-iEt} 
\begin{pmatrix}
	\phi
	\\
	\chi
\end{pmatrix}
\ee
where $\phi$ and $\chi$ are the two-component spinors. Inserting Eq.(\ref{equ_genesolu}) into Eq.(\ref{equ_diraequ}), where chiral representations of the $\gamma$-matrices are used,  we obtain the coupled equations for $\phi$ and $\chi$:
\begin{align}
	\label{equp}(M - T \sigma^3 )\phi - (E+i \vec{\sigma}\cdot \vec{D})\chi   & = 0,\\
	\label{eqdo}(E - i \vec{\sigma}\cdot \vec{D})\phi - (M - T \sigma^3 )\chi & = 0.
\end{align}
Eliminating $\chi$ from Eq.(\ref{equp}) and Eq.(\ref{eqdo}) then the equation for $\phi$ is obtained:
\be
\label{equ_Aphi}
\hat{A}\phi={\Big \{ }(M - T \sigma^3)(M^2-T^2)- (E + i \vec{\sigma}\cdot \vec{D})(M +T \sigma^3 )(E - i \vec{\sigma}\cdot \vec{D}) {\Big \}} \phi=0.
\ee
It is obvious that $\phi$ is a two-component spinor while $\hat{A}$ is a $2 \times 2$ matrix, and the elements of $\hat{A}$ are listed as following:
\be
\label{equ_A}
\begin{split}
	\hat{A}_{12} &=2T (E+iD_3) (iD_1+D_2) ,\\
	\hat{A}_{21} &=-2T (E - iD_3) (iD_1 - D_2),\\
	\hat{A}_{11} &=(M + T)(M - T )^2 -(M+T)(E^2+D_3^2)+(M - T) (iD_1+D_2)(iD_1-D_2),\\
	\hat{A}_{22} &=(M - T )(M + T )^2 - (M-T)(E^2+D_3^2)+(M + T) (iD_1-D_2)(iD_1+D_2).
\end{split}
\ee
To determine the form of $\phi$, let's first consider function $f_{k}(x)$ defined as below:
\be
f_k(x)=c_k \text{e}^{-\frac{1}{2}(\frac{x}{l} - p_y l)^2} H_k(\frac{x}{l} - p_y l) \text{e}^{i(\xi_f p_y y+p_z z)},
\ee
where $H_k(x)$ is the Hermite polynomials, $c_k$ is the normalized constant, with $l=1/\sqrt{|q B|}$ , $\xi_f=\text{sgn}(q)$, and we set $f_{-1}=0$. With a little effort the following relations for $q>0$ can be obtained:
\begin{align}
	\label{equ_cre}(iD_1+D_2)f_k &= -i\sqrt{|q B|} c_{k,k+1} f_{k+1},\\
	\label{equ_ann}(iD_1-D_2)f_k &=2ik \sqrt{|q B|} c_{k,k-1} f_{k-1},	
\end{align}
where $c_{n,m}=c_n/c_m$. In fact, the operators $iD_1+D_2$ and $iD_1-D_2$ are creation and annihilation operator, respectively. And then the general form of $\phi$ is straightforward:
\be\label{equ_phi}
\phi_{k}(x)=
\begin{pmatrix}
	f_k(x)
	\\
	f_{k-1}(x)
\end{pmatrix}.
\ee
Now insert Eq.(\ref{equ_A}) and Eq.(\ref{equ_phi}) into Eq.(\ref{equ_Aphi}) and we obtain two de-coupled equations:
\be
\begin{split}
	& \Big \{(M + T ) (M - T )^2 - (M+T)(E^2-p_z^2) +  2k (M - T)|q B| \Big \} f_k\\
	& \quad \quad \quad \quad \quad \quad \quad \quad \quad \quad \quad \quad \quad \quad \quad \quad -2iT (E - p_z)  \sqrt{|q B|} c_{k-1,k} f_{k}=0,
\end{split}
\ee
\be
\begin{split}	
	& \Big \{ (M-T) (M + T )^2-(M-T)(E^2-p_z^2)+2k (M + T)|q B| \Big \} f_{k-1} \\
	&\quad \quad \quad \quad \quad \quad \quad \quad \quad \quad \quad \quad \quad \quad \quad - 2i T (E + p_z)  \cdot 2k\sqrt{|q B|} c_{k,k-1} f_{k-1} =0.
\end{split}
\ee
Combining above two equations with relations Eq.({\ref{equ_cre}}) and Eq.(\ref{equ_ann}), the dispersion relation for positive-charged fermion with spin-$s$ is obtained as:
\be
E_{k}^2 = 
\begin{cases} 
	p_z^2+\{\sqrt{M^2+2k |q B|}-s T \}^2, &\mbox{if } k \ge 1 \\
	&   \\
	p_z^2+(M-T)^2, & \mbox{if } k=0 
\end{cases}
\label{equ_disp_pos}
\ee
A similar relation for negative charged fermion can be obtained, however, we won't repeat here. Finally, taking $T= \kappa q B$, the dispersion relation for fermions with charge $q$, spin-$s$ and anomalous magnetic moment $\kappa$ in the presence of magnetic field is:
\be
\label{equ_disp}
E_{k}^2=p_z^2+\{\sqrt{M^2+(2k+1-s\xi)|q B|}-s \kappa q B\}^2,
\ee
where $ s=\pm 1$ is for spin-up and spin-down,respectively, and $\xi=\text{sgn}(q)$. From this dispersion relation, we can see that Lowest Landau Level (LLL) 
for both spin-up/down positive/negative charged quarks have the form of   
\begin{eqnarray}
E_{0}^2 &=& p_z^2 + (M - \kappa |q_f| B)^2,~~~~ u^{\uparrow},\bar{d}^{\uparrow}  \nonumber \\
E_{0}^2 &=&p_z^2 + (M - \kappa|q_f| B)^2, ~~~~ \bar{u}^{\downarrow},d^{\downarrow}  \nonumber \\
E_{0}^2 &=&p_z^2 + (\sqrt{M^2+2|q_fB|} + \kappa |q_f| B)^2, ~~~~ \bar{u}^{\uparrow},d^{\uparrow}  \nonumber \\
E_{0}^2 &=&p_z^2 + (\sqrt{M^2+2|q_fB|} + \kappa |q_f| B)^2. ~~~~u^{\downarrow},\bar{d}^{\downarrow} 
\label{equ:LLL}
\end{eqnarray}
Therefore, for two-flavor quark system, for $\kappa>0$, the lowest energy is occupied by spin-up positive-charged fermions $u^{\uparrow},\bar{d}^{\uparrow}$ and spin-down negative-charged fermions $\bar{u}^{\downarrow},d^{\downarrow}$. For higher excitations, for either positive or negative charged fermions, the energy spectrum exhibits a Zeeman splitting($s=\pm1$), which can be seen clearly in Fig.(\ref{fig:AMM-dis}). 

\subsection{Inverse Magnetic Catalysis with AMM}
From the dispersion relation Eq.(\ref{equ:LLL}) in the LLL, e.g., $E_{0}^2 = p_z^2 + (M - \kappa |q_f| B)^2$, we can see that AMM reduces the dynamical quark mass. We numerically investigate the effect of AMM on the dynamical quark mass, and following Ref.\cite{Chaudhuri:2019lbw,Chaudhuri:2020lga,Fayazbakhsh:2014mca,Klevansky:1992qe} we obtain the one-loop level effective potential at zero baryon chemical potential and finite temperature:
\be
\label{equ:gapequT}
\begin{split}
	\Omega=\frac{(M-m_0)^2}{4G_S} & - N_c\sum_{f}\frac{|q_{f}B|}{2\pi}\sum_{n}\sum_{s=\pm 1}\int \frac{dp_z}{2\pi}E_{n,f,s}\\
	& -  2N_cT\sum_{f}\frac{|q_{f}B|}{2\pi}\sum_{n}\sum_{s=\pm 1}\int \frac{dp_z}{2\pi}\ln(1+\text{e}^{-\frac{E_{n,f,s}}{T}}).
\end{split}
\ee
here only scalar channel is considered for the study of the dynamical quark mass, which can be obtained by solving the gap equation:
\begin{equation}
	\frac{\partial \Omega}{\partial M}  =0,
\end{equation}
and in explicit:
\begin{equation}
\frac{M-m_0}{2G_S}=N_c \sum_{f}\frac{|q_fB|}{2\pi}\sum_{n}\sum_{s=\pm 1}\int \frac{dp_z}{2\pi} {\Big \{ }1-2(1+\text{e}^{\frac{E_{n,f,s}}{T}})^{-1} {\Big \} }\frac{M}{E_{n,f,s}}(1-\frac{s\kappa q_f B}{M_n}),
\end{equation}
where $M=m_0+\sigma$ is quark's dynamical mass with
\begin{equation}
\sigma=2G_SN_c \sum_{f}\frac{|q_fB|}{2\pi}\sum_{n}\sum_{s\pm 1}\int \frac{dp_z}{2\pi} {\Big \{ }1-2(1+\text{e}^{\frac{E_{n,f,s}}{T}})^{-1} {\Big \} }\frac{M}{E_{n,f,s}}(1-\frac{s\kappa q_f B}{M_n}),
\label{eq:sigma}
\end{equation}
where $M_n=\sqrt{M^2+(2n+1-s\xi_f)|q_fB|}-s \kappa q_f B$. In strong magnetic field region, we can take LLL approximation then the gap equation 
at zero temperature becomes:
\begin{equation}
\begin{split}
M & =m_0+2G_S N_c\sum_f\frac{|q_fB|}{2\pi}\int\frac{dp_z}{2\pi}\frac{1}{E_0}(1-\frac{\kappa|q_fB|}{M}) \\
& =m_0+ 2G_S N_c\sum_f\frac{|q_fB|}{2\pi}\int\frac{dp_z}{2\pi}\frac{1}{E}(1-\lambda \kappa+(\kappa^2)),
\end{split}
\end{equation}
where $E=\sqrt{p_z^2+M^2}$ and $\lambda=\frac{p_z^2}{M(M^2+p_z^2)}$. In the second line we assume a small $\kappa$ expansion, and it's clear that a none-zero $\kappa$ reduces the quark mass $M$ in the strong magnetic field region. 

\begin{figure}[h!]
	\centering 
	\includegraphics[width=8cm]{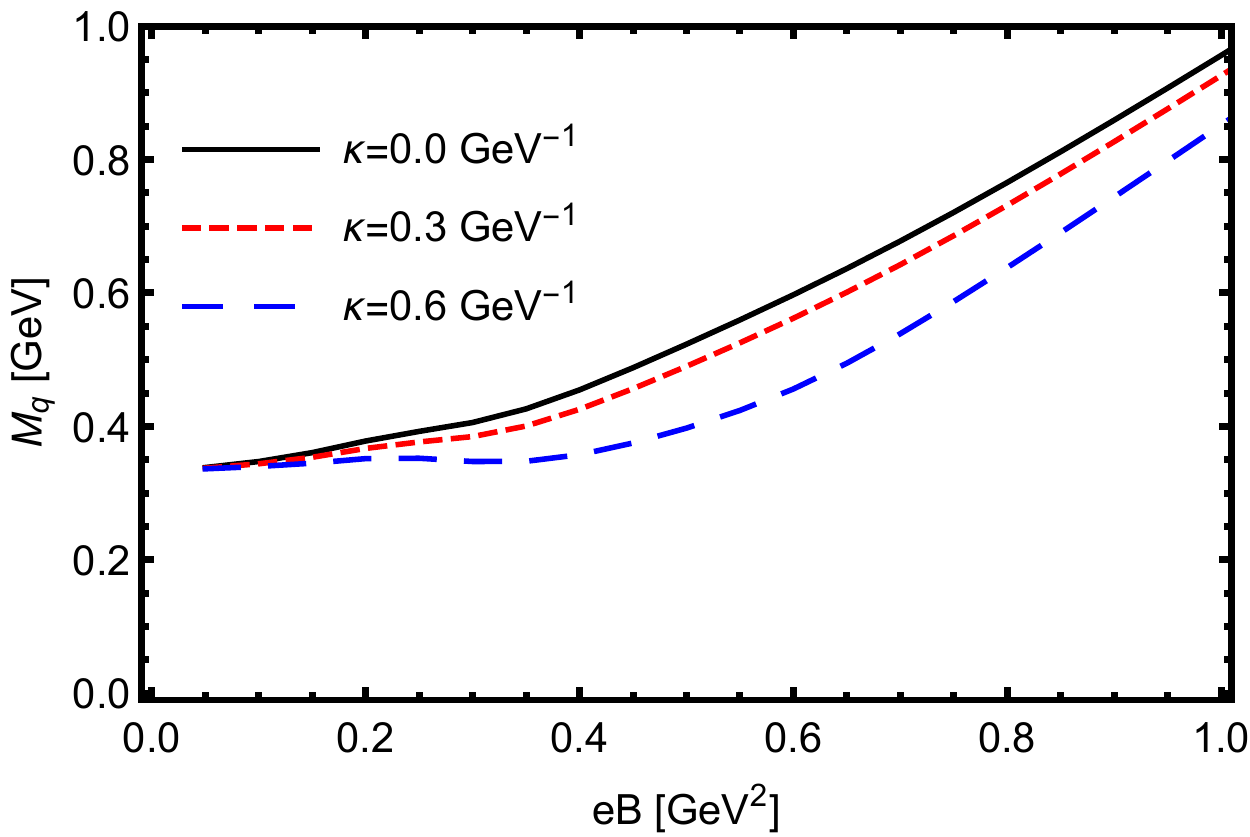}
	\includegraphics[width=8cm]{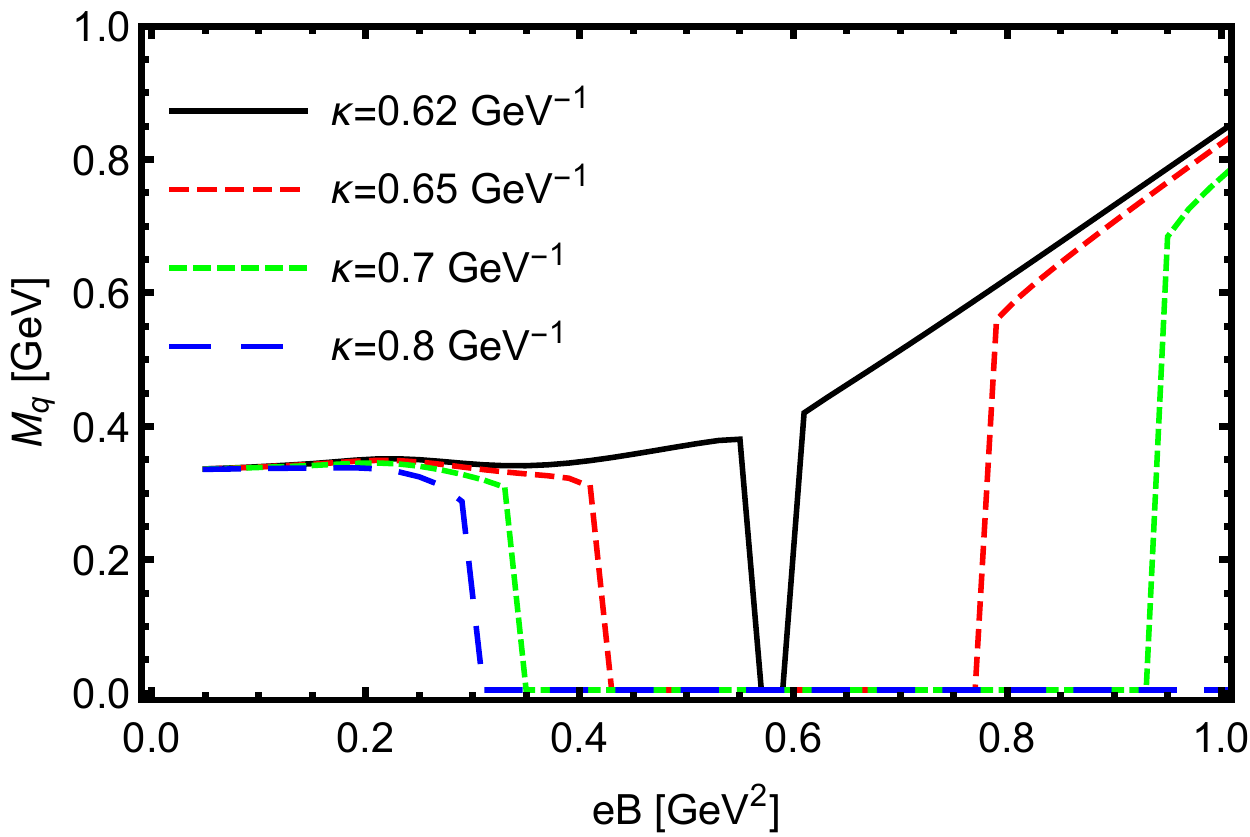}
	\caption{Dynamical quark mass as a function of magnetic field with different constant $\kappa$ at zero temperature.}
	\label{fig:quarkmass}
\end{figure}

\begin{figure}
	\includegraphics[width=10cm]{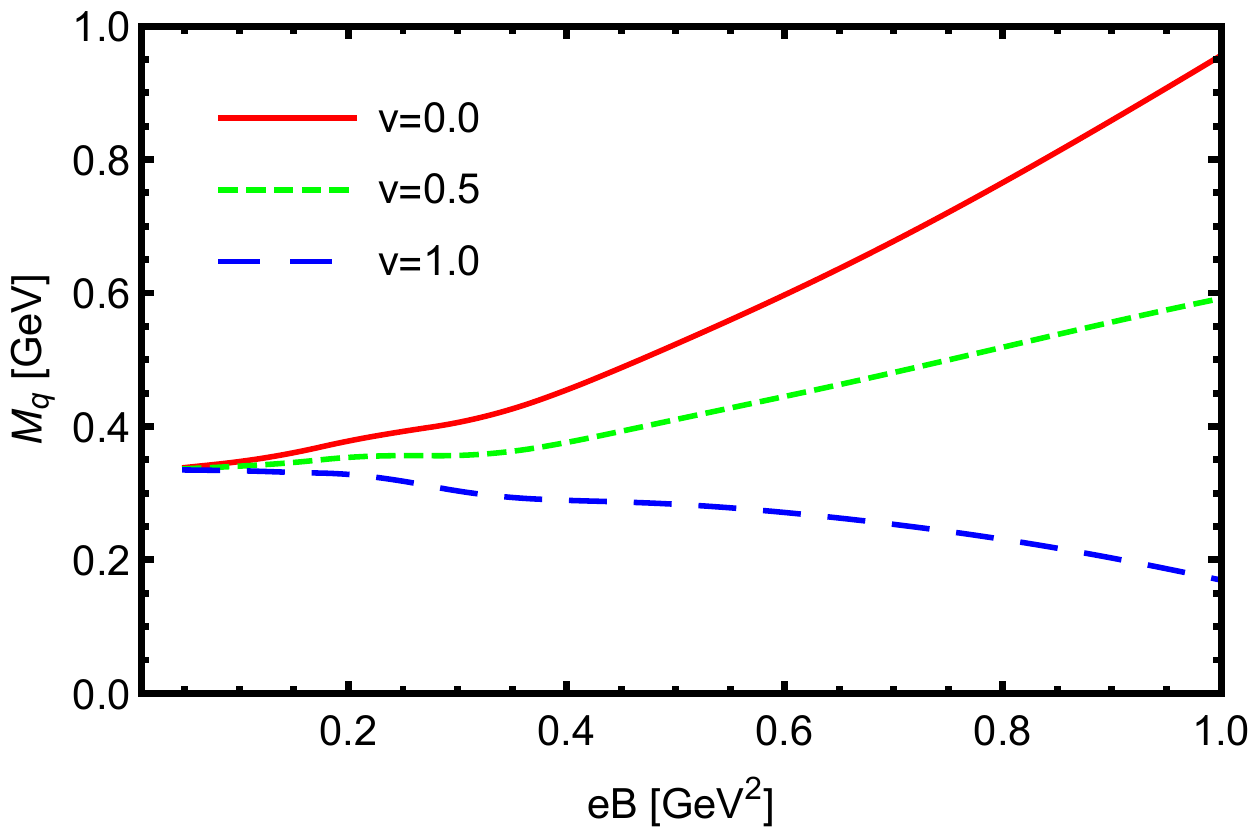}\hspace{1pt}
	\caption{Dynamical quark mass as a function of magnetic field with $\kappa=v\sigma$ at zero temperature.}
	\label{fig:quarkmass_eB_v}
\end{figure}

The numerical results of the dynamical quark mass as a function of the magnetic field with AMM is shown in Fig.(\ref{fig:quarkmass}) and Fig.(\ref{fig:quarkmass_eB_v}), where $\kappa$ is treated as a constant and $\kappa \sim \sigma$ with $\sigma$ solved from Eq.(\ref{eq:sigma}), respectively. 

\begin{figure}[h!]
	\centering 
	\subfloat[]{\includegraphics[width=8cm]{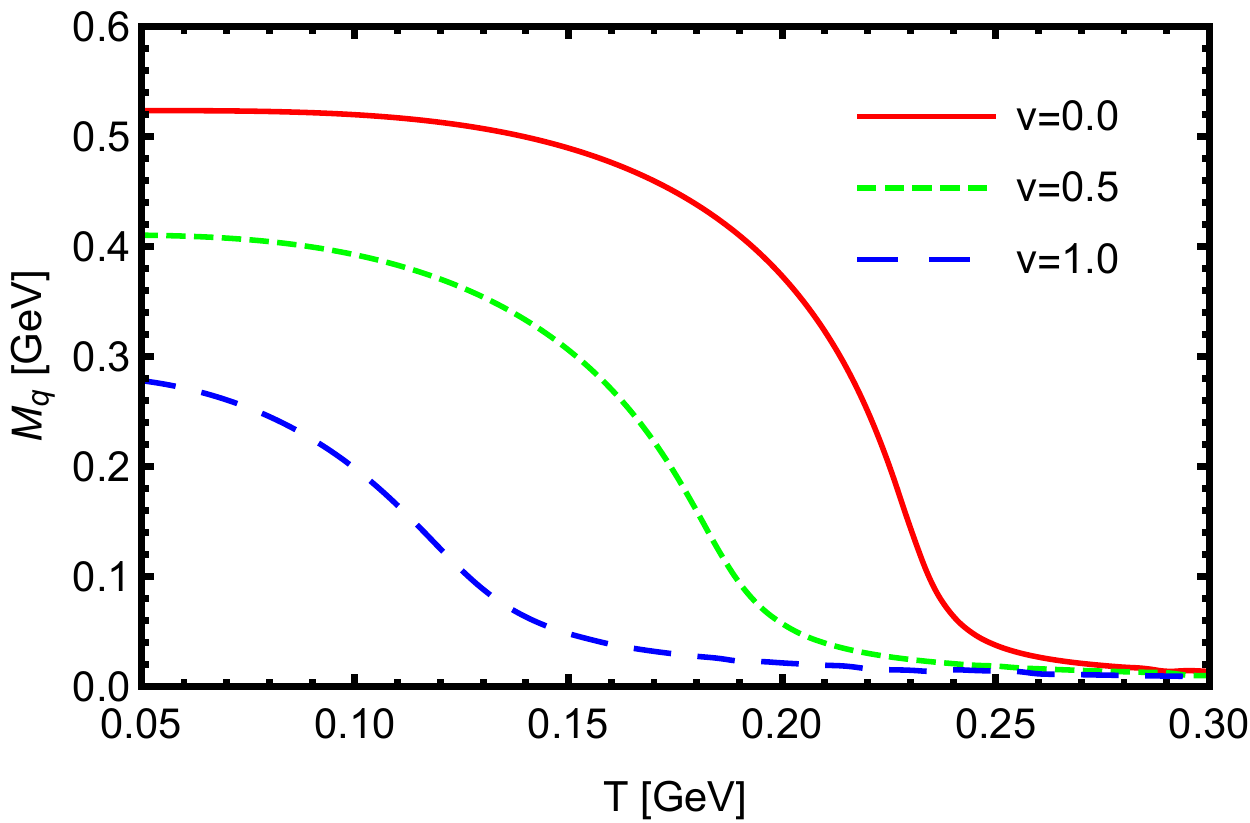}\label{subfig:Mq_T_v}}
	\subfloat[]{\includegraphics[width=8cm]{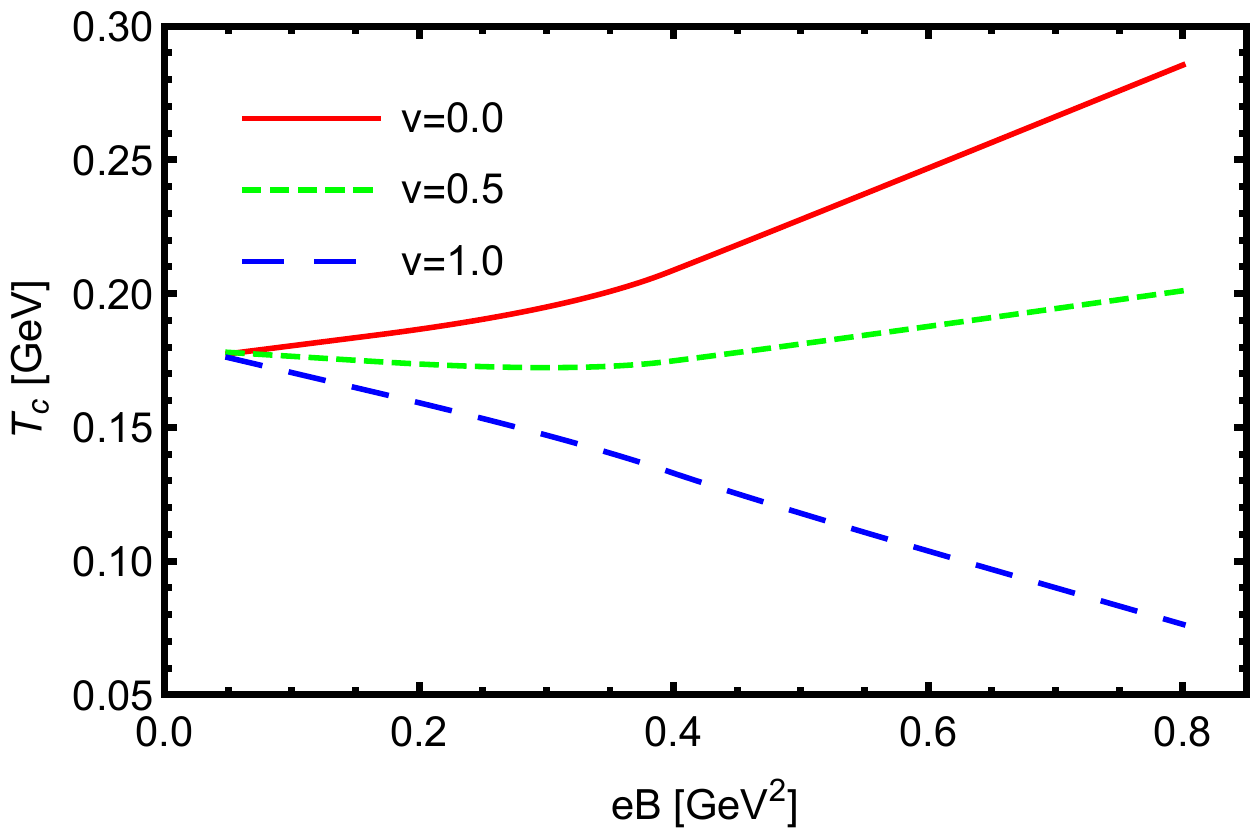}\label{subfig:Tc_eB_v}}
	\caption{(a): Dynamical quark mass $M_q$ as a function of temperature with $\kappa=v \sigma$ at a fixed magnetic field $eB=0.5\text{GeV}^2$. (b): critical temperature $T_c$ of chiral transition aas a function of magnetic field. With proper $v$, Inverse Magnetic Catalysis can be obtained(e.g., blue dashed line).}
	\label{fig:IMC}
\end{figure}
1) We firstly treat $\kappa$ as a free constant and calculate the dynamical quark mass as a function of the magnetic field at zero temperature with different $\kappa$, and numerical results are shown in Fig.(\ref{fig:quarkmass}). It can be observed clearly that the magnetic catalysis (MC) effect is competing with the mass reducing effect from the the AMM. For small $\kappa \lesssim 0.6 \text{GeV}^{-1}$, the dynamical quark mass $M_q$ increases with magnetic field, which is known as Magnetic Catalysis, but the AMM explicitly reduces dynamical quark mass as shown in the left figure of Fig.(\ref{fig:quarkmass}), which is consistent with the dispersion relation of quark with AMM as shown in Eq.(\ref{equ:LLL}). For larger $\kappa$, as shown in the right figure of Fig.(\ref{fig:quarkmass}), the behavior of dynamical quark mass becomes complicated. For $\kappa=0.62 \text{GeV}^{-1}$, $M_q$ still increases with magnetic field but drops to zero and then jumps to none zero at a narrow region of magnetic field around $eB_c\sim 0.6\text{GeV}^2$. As $\kappa$ increases, the region that $M_q=0$ becomes larger and larger, the left edge of which decreases while the right edge increases. And when $\kappa=0.8 \text{GeV}^{-1}$ as the blue dashed line shown in Fig.(\ref{fig:quarkmass}), the dynamical quark mass slightly decreases as magnetic field increases then drops to zero at $eB_c \sim 0.3 \text{GeV}^{2}$ and keeps zero in the rest region we considered $eB < 1\text{GeV}^{2}$.

2) Then we take $\kappa$ proportional to be to the quark condensate $\kappa= v \sigma$ with $v$ the ratio and $\sigma$ solved from Eq.(\ref{eq:sigma}). The corresponding result for dynamical quark mass at zero temperature is shown in Fig.(\ref{fig:quarkmass_eB_v}). We can see that when the ratio $v$ is small, as shown by the red solid line and green dashed line, $M_q$ increases with magnetic field monotonically, the MC effect overcomes the mass reducing effect by the AMM. When $v$ is large enough, the mass reducing effect induced by the AMM dominates thus one can see that the dynamical quark mass decreases smoothly with the increasing magnetic field, no "jump" behavior shows up as the case of constant $\kappa$. Then we also consider dynamical quark mass at finite temperature, under a fixed magnetic field $eB=0.5\text{GeV}^2$, as shown in Fig.(\ref{subfig:Mq_T_v}), it's clear that larger $v$ can not only reduce the dynamical quark mass, also reduce the critical temperature $T_c$, which is determined through $T_c=(-\partial M_q/\partial T)$. $T_c$ as a function of the magnetic field under different $v$ is plotted in Fig.(\ref{subfig:Tc_eB_v}): for small $v$, conventional Magnetic Catalysis shows up, i.e., $T_c$ increases with the magnetic field, however, when $v$ is large, for example, $v=1$ as shown by the blue dashed line, $T_c$ decreases with the magnetic field, which indicates the Inverse Magnetic Catalysis.
\hskip 2cm

\subsection{Magnetic Susceptibility with AMM}

\begin{figure}[h!]
	\centering
	\includegraphics[width=10cm]{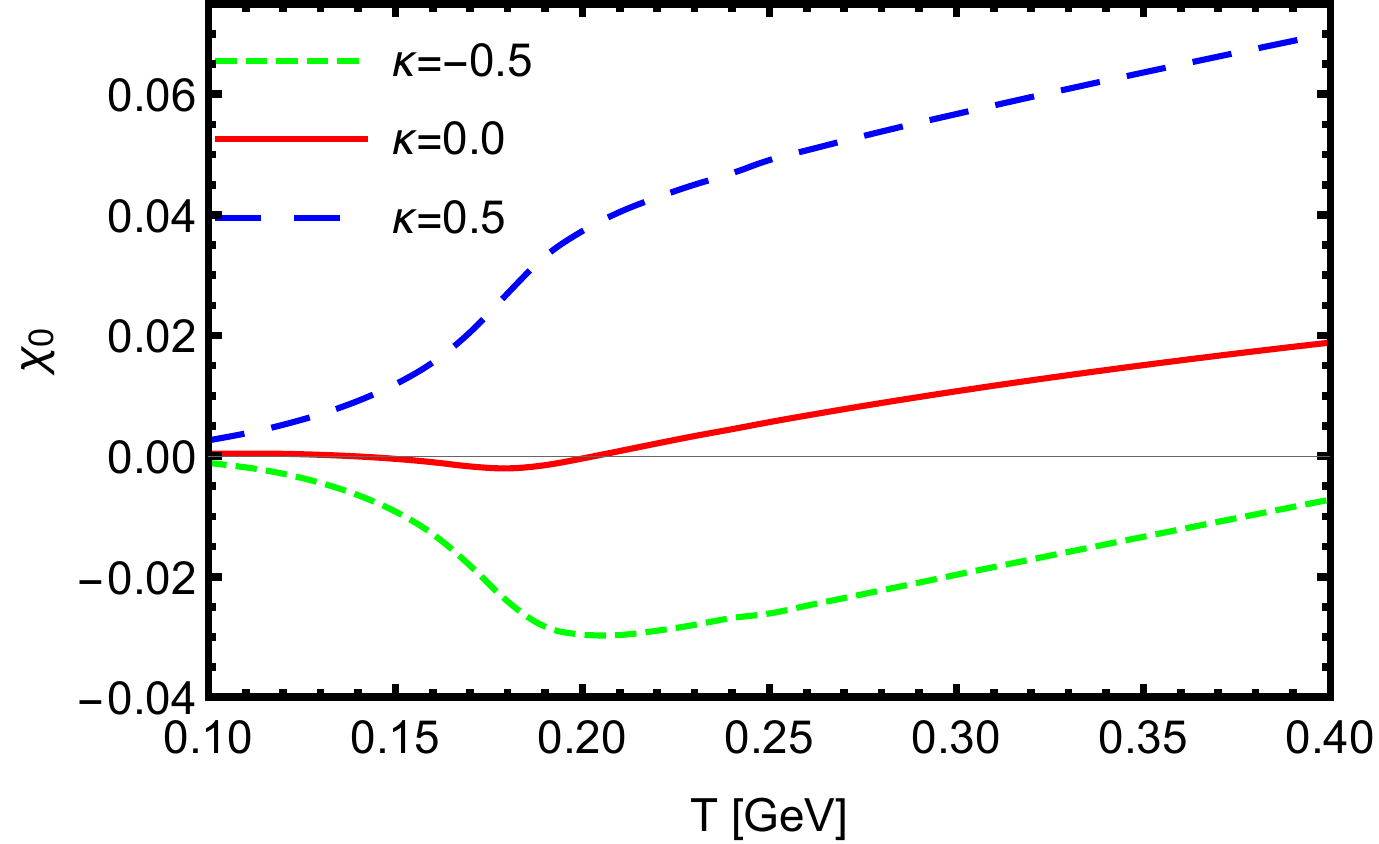}
	\caption{Magnetic susceptibility with different constant $\kappa$. A soft cutoff is applied during the numerical calculation.}	
	\label{fig:sus}
\end{figure}

Recently, lattice calculations show more interesting and novel properties of magnetized QCD matter: magnetized matter exhibits diamagnetism (negative susceptibility) at low temperature and paramagnetism (positive susceptibility) at high temperature~\cite{Bali:2012jv,Bali:2020bcn}. From the above dispersion relation, we can see that the AMM of quarks causes Zeeman splitting in the dispersion relation thus changes the magnetism properties. We show the numerical results for magnetic susceptibility induced by the AMM. The magnetic susceptibility is defined as:
\be
\chi=-\frac{\partial^2 \Omega}{\partial (eB)^2}|_{eB=0},
\ee
and we define  $\chi_0(T)=\chi(T)-\chi(T=0)$, the numerical result of which is shown in Fig.(\ref{fig:sus}), where soft cutoff is applied. We can see that at high temperature, there is no doubt that the magnetized matter shows paramagnetism with $\chi_0>0$. However, at low temperature, a negative $\kappa$ gives negative susceptibility, i.e, which indicates the diamagnetism of magnetized QCD matter, while a positive $\kappa$ gives positive susceptibility, i.e, the paramagnetism of magnetized QCD matter. Therefore, the diamagnetism property cannot be understood by considering the AMM of quarks.

\section{MESON SPECTRA WITH AMM}
\label{sec-meson}
Next, we investigate meson spectra with the AMM. In the NJL model, mesons are regarded as  $q \bar q $ bound states or resonances, which can be obtained from the quark-antiquark scattering amplitude~\cite{Klevansky:1992qe,He:1997gn,Rehberg:1995nr}. As shown in Fig.(\ref{fig:mesonRPA}), The full propagator of meson can be expressed to leading order in $1/N_c$ as an infinite sum of quark-loop chains under the random phase approximation (RPA). Following the procedure given in Ref.\cite{Miransky:2015ava}, the quark propagator in the Landau level representation is given by
\begin{equation}
S(x,y)=\text{e}^{i\Phi_f(x,y)}\int \frac{d^4 q}{(2\pi)^4}\text{e}^{-i(x-y)q}\widetilde{S}(q),
\end{equation}
where the Schwinger phase $\Phi_f(x,y)=q_f (x^1+y^1)(x^2-y^2)/2$ breaks the translation invariant while $\widetilde{S}(q)$ is translation-invariant, which takes the form of :
\bea
\label{equ:quarkpropa}
\widetilde{S}_f(k)&=& i\exp\left(-\frac{\mathbf{k}_{\bot}^2}{|q_f B|}\right)\sum_{n=0}^{\infty}(-1)^n\frac{D_n(q_f B,k)F_{n}(q_f B,k)}{A_{n}(q_f B,k)},
\eea
where:
\begin{equation}
F_{n}(q_f B,k)=(\kappa q_fB- k^0 \gamma^3 \gamma^5+k^3 \gamma^0 \gamma^5)^2-M^2-2 n |q_fB|,
\end{equation}
and 
\bea\label{Dn}
D_n(q_f B,k)&=&(k^0\gamma^0-k^3\gamma^3+M+\kappa q_f B \sigma^{12})\Big[(1-i\gamma^1\gamma^2\xi_{f})L_n\left(2\frac{\mathbf{k}_{\bot}^2}{|q_f B|}\right)-(1+i\gamma^1\gamma^2\xi_{f})L_{n-1}\left(2\frac{\mathbf{k}^2_{\bot}}{|q_f B|}\right)\Big]\nonumber\\
& + &4(k^1\gamma^1+k^2\gamma^2)L_{n-1}^1\left(2\frac{\mathbf{k}^2_{\bot}}{|q_f B|}\right),\nonumber\\
\eea
where $\xi_{f}=\text{sign}(q_fB)$, $L_n^{\alpha}$ are the generalized Laguerre polynomials and $L_n=L_n^0$. And the Denominator:
\begin{equation}
A_{n}(q_f B,k)=\Pi_{s=\pm1}\Big{\{}\Big{(}\kappa q_fB+s\sqrt{(k^0)^2-(k^3)^2} \Big{)}^2-M^2-2n|q_fB|\Big{\}}^2.
\end{equation}
With RPA approximation, the composite $\pi$ propagator is written as:
\begin{figure}[h!]
	\centering
	\includegraphics[width=15cm]{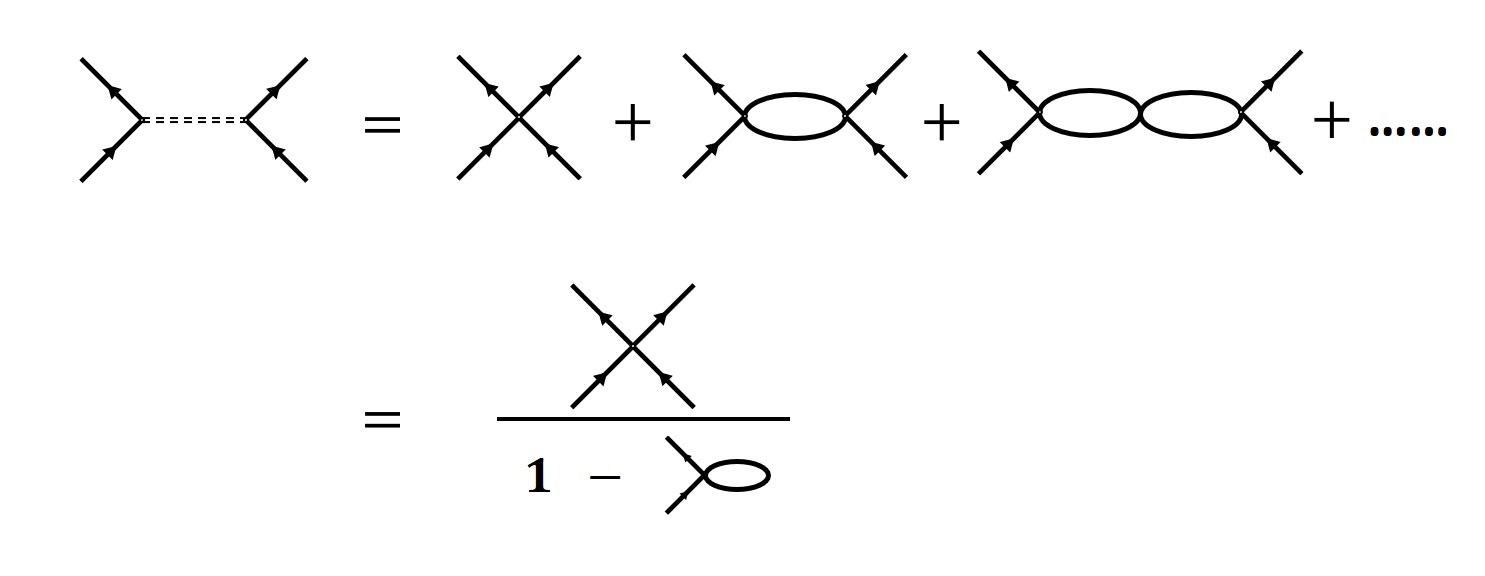}
	\caption{Meson propagator under the Random Phase Approximation in the NJL model.}
	\label{fig:mesonRPA}
\end{figure}
\begin{equation}
	D_{\pi}(q^2)=\frac{2G_{S}}{1-2G_{S}\Pi_{\pi}(q^2)},
\end{equation}
where $\Pi_{\pi}$ is the one loop polarization function for pion:
\begin{eqnarray}
	\label{pola_pi}
	\Pi_{\pi}(q^2)  =  i \int \frac{d^4k}{(2\pi)^4} \text{Tr}[\tau^a i\gamma^5 \widetilde{S}(k)\tau^b i\gamma^5 \widetilde{S}(p)],
\end{eqnarray}
with $q=k-p$. For neutral pion, the Pauli matrices take $\tau^a=\tau^3,\tau^b=\tau^3$  while for charged pion they take $\tau^a=\tau^{\pm},\tau^b=\tau^{\mp}$ with definition:
\be
\tau^{\pm}=\frac{1}{\sqrt{2}}(\tau^1\mp i\tau^{2}).
\ee
And $\widetilde{S}(q)$ is the translation-invariant part of quark propagator in momentum space and the explicit expression is given in Eq.(\ref{equ:quarkpropa}). From the pole of propagator $D_{\pi}(q^2)$, the corresponding meson mass can be obtained by solving:
\begin{equation}
1-2 G_S \Pi_{\pi}(q^2=m_{\pi}^2)=0.
\end{equation}
It's similar for the vector meson rho with corresponding one loop polarization function $\Pi_{V}^{\mu\nu}$:
\begin{eqnarray}
	\label{pola_rho}
	\Pi_{V}^{\mu\nu}(q^2)  =  i \int \frac{d^4k}{(2\pi)^4} \text{Tr}[\tau^a \gamma^{\mu} \widetilde{S}(k)\tau^b \gamma^{\nu} \widetilde{S}(p)],
\end{eqnarray}
and it can be easily proved that the polarization function for charged rho has following structure:
\be
\Pi^{\mu\nu}_{\pm}=
\begin{pmatrix}
	\Pi_{\pm}^{00} & 0  & 0 & 0 \\
	0 & \Pi_{\pm}^{11} & \Pi_{\pm}^{12} & 0 \\
	0 & \Pi_{\pm}^{21} & \Pi_{\pm}^{22} & 0 \\
	0 & 0  & 0 & \Pi_{\pm}^{33} \\
\end{pmatrix},
\ee
where $\Pi_{\pm}^{11}=\Pi_{\pm}^{22}$ and $\Pi_{\pm}^{12}=-\Pi_{\pm}^{21}$. One can also decompose it into four parts with respect to z-component of spin, e.g., $s_z$, in the rest frame~\cite{Liu:2014uwa}:
\be
	\Pi^{\mu\nu}_{\pm}(q)=\Pi^{s_z=+1}_{\pm} \epsilon^{\star,\mu}_1\epsilon^{\nu}_1+\Pi^{s_z=-1}_{\pm} \epsilon^{\star,\mu}_2\epsilon^{\nu}_2+\Pi^{s_z=0}_{\pm} b^{\mu}b^{\nu}+\Pi^{u}_{\pm}u^{\mu}u^{\nu},
\ee
where $u^{\mu}=(1,0,0,0)$ is the four momentum in the rest frame and spin projection operator are introduced:
\be 
	\epsilon^{\mu}_1=\frac{1}{\sqrt{2}}(0,1,i,0), \quad \epsilon^{\mu}_2=\frac{1}{\sqrt{2}}(0,1,-i,0), \quad
	b^{\mu}=(0,0,0,1).
\ee
It's worthy to point out that the last term $\Pi^{u}_{\pm}$ corresponds to un-physical component of charged rho polarization function. As a consequence, the charged rho propagator can be written as:
\be
	D^{\mu\nu}_{\pm}=D^{s_z=+1}_{\pm} \epsilon^{\star,\mu}_1\epsilon^{\nu}_1+D^{s_z=-1}_{\pm} \epsilon^{\star,\mu}_2\epsilon^{\nu}_2+D^{s_z=0}_{\pm} b^{\mu}b^{\nu}+D^{u}_{\pm}u^{\mu}u^{\nu},
\ee
then we obtain propagator for each component:
\be
	D^{s_z}_{\pm}(q)=\frac{2G_V}{1+2G_V \Pi^{s_z}_{\pm}(q)},
\ee
and the mass for each component can be obtained by solving following equations:
\bea
	1+2G_V \Pi_{\pm}^{s_z=+1}(q^2=m_{\rho^{\pm},s_z=+1}) & = &	1+2G_V (\Pi_{\pm}^{11}-i\Pi_{\pm}^{12})=0, \\
	1+2G_V \Pi_{\pm}^{s_z=-1}(q^2=m_{\rho^{\pm},s_z=-1}) & = & 1+2G_V (\Pi_{\pm}^{11}+i\Pi_{\pm}^{12})=0, \\
	1+2G_V \Pi_{\pm}^{s_z=0}(q^2=m_{\rho^{\pm},s_z=0}) & = &1+2G_V \Pi_{\pm}^{33}=0.
\eea
For neutral rho, the one loop polarization function has structure:
\be
\Pi^{\mu\nu}_{0}=
\begin{pmatrix}
	\Pi_{0}^{00} & 0  & 0 & 0 \\
	0 & \Pi_{0}^{11} & 0 & 0 \\
	0 & 0 & \Pi_{0}^{22} & 0 \\
	0 & 0  & 0 & \Pi_{0}^{33} \\
\end{pmatrix},
\ee
where $\Pi_{0}^{11}=\Pi_{0}^{22}$, and now the gap equations for neutral rho with $s_z=\pm1,0$ are:
\bea
1+2G_V \Pi_{0}^{s_z=\pm1}(q^2=m_{\rho^{0},s_z=\pm1}) & = &	1+2G_V \Pi_{0}^{11}=0, \\
1+2G_V \Pi_{0}^{s_z=0}(q^2=m_{\rho^{0},s_z=0}) & = &1+2G_V \Pi_{0}^{33}=0. 
\eea

For neutral pion and rho, the Schwinger phase in quark-antiquark loop cancels out while for charged pion and rho they not, for the Schwinger phase led to more complicated calculation,  in the present paper, we ignore the Schwinger phase and only consider the translation invariant part $\widetilde{S}$.

\begin{figure}[h!]
	\centering
	\subfloat[]{\includegraphics[width=8cm]{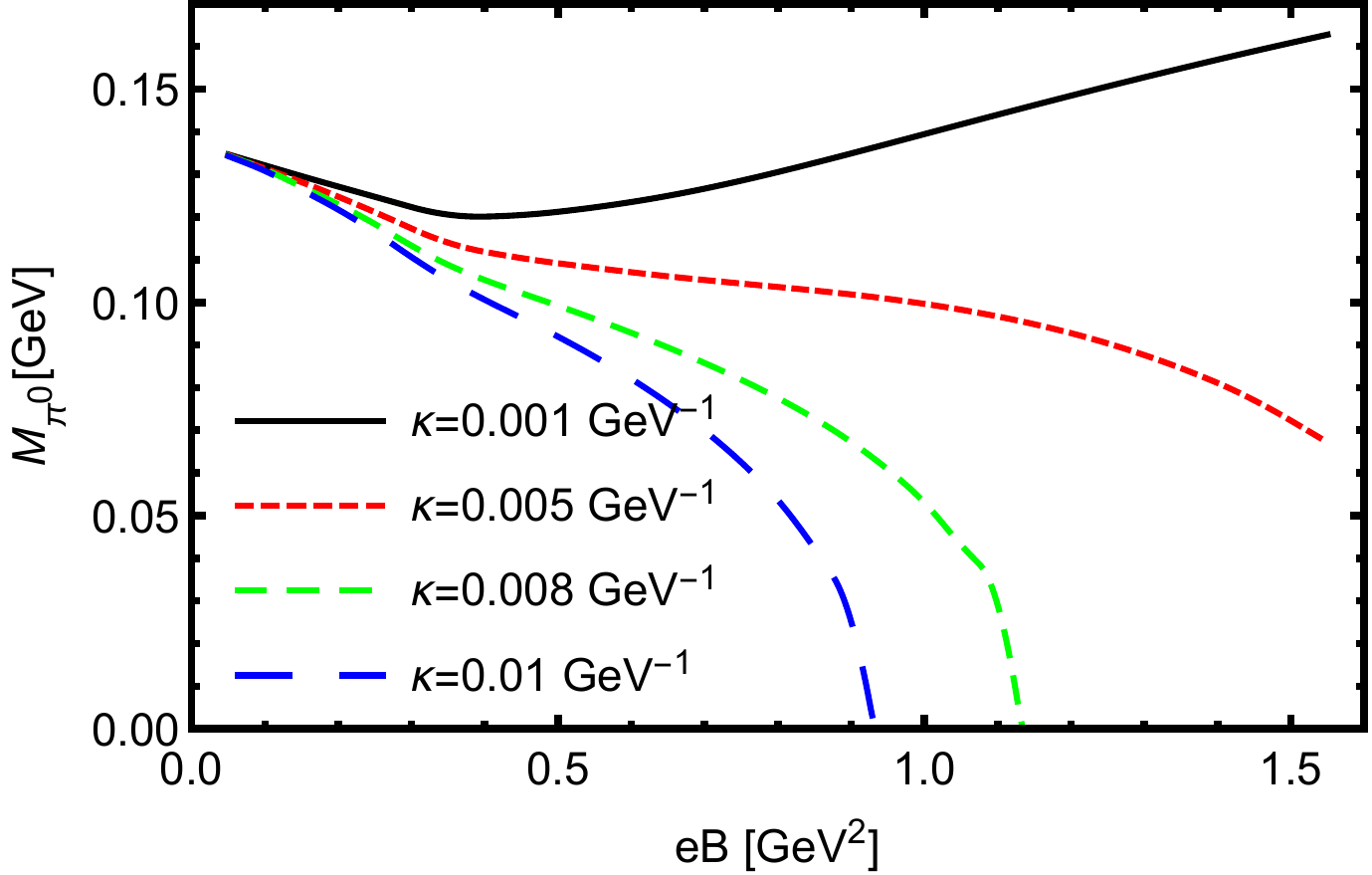}\label{subfig:neutralpionmass}}
	\subfloat[]{\includegraphics[width=8cm]{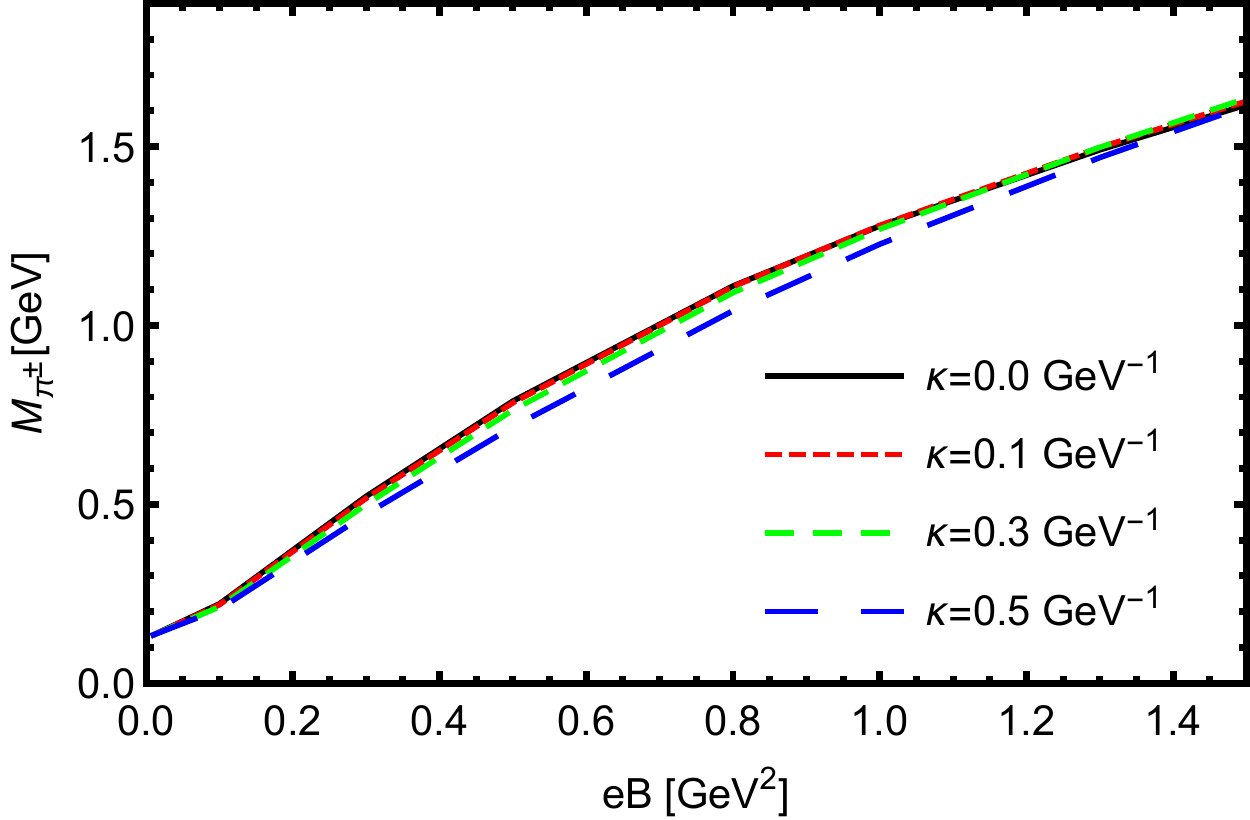}\label{subfig:chargedpionmass}}
	\caption{(a)Neutral and (b)charged pion mass as a function of the magnetic field with different constant $\kappa$.}
	\label{fig:pionmass}
\end{figure}

We present numerical results for the mass of pion and rho at zero temperature, where a soft cut-off is applied. The neutral and charged pion mass as a function of magnetic field are shown in Fig.(\ref{fig:pionmass}). It is found that The presence of AMM reduces neutral pion mass $M_{\pi^0}$ significantly, and $M_{\pi^0}$ is sensitive to the AMM of quarks, as shown in Fig.(\ref{subfig:neutralpionmass}).  In the case without AMM, neutral pion mass decreases slightly as magnetic field increases then it increases after a inflection point. When an appropriate AMM is applied, the previous inflection point disappears, $M_{\pi^0}$ continuously decreases and reaches zero at a critical magnetic field point $eB_c$. And as $\kappa$ increases, $eB_c$ decreases, which indicates a quick drop of $M_{\pi^0}$ as magnetic field increases: when $\kappa=0.005\text{GeV}^{-1}$, $eB_c$ is larger than $1.5\text{GeV}^{2}$, and for $\kappa=0.008\text{GeV}^{-1}$,  $eB_c\sim 1.15\text{GeV}^{2}$ while $eB_c\sim 0.95\text{GeV}^{2}$ for $\kappa=0.01\text{GeV}^{-1}$.  It has been studied in many papers~\cite{Coppola:2018vkw,Coppola:2019uyr,Mao:2018dqe,Wang:2017vtn} that magnetic field increases charged pion mass $M_{\pi^{\pm}}$, and as shown in Fig.(\ref{subfig:chargedpionmass}), $M_{\pi^{\pm}}$ increases with magnetic field in both zero and non-zero AMM cases. Similar to neutral pion, AMM also reduces charged pion mass, however, the modification from AMM is slight enough to be ignored. Besides, comparing the results of neutral pion with the charged pion, it's obvious that they exhibit different sensitivities to the AMM:  A very small $\kappa$ of AMM can change the behavior of neutral mass behavior in the region of $eB>0.4\text{GeV}^2$(dashed blue line shown in Fig.(\ref{subfig:neutralpionmass})) from increasing with magnetic field to decreasing, while the behavior of charged pion mass hardly changes even up to $\kappa=0.5\text{GeV}^{-1}$(dashed blue line shown in Fig.(\ref{subfig:chargedpionmass})) at the range of magnetic field $0<eB<1.5\text{GeV}^2$.

\begin{figure}[h!]
	\centering
	\subfloat[]{\includegraphics[width=8cm]{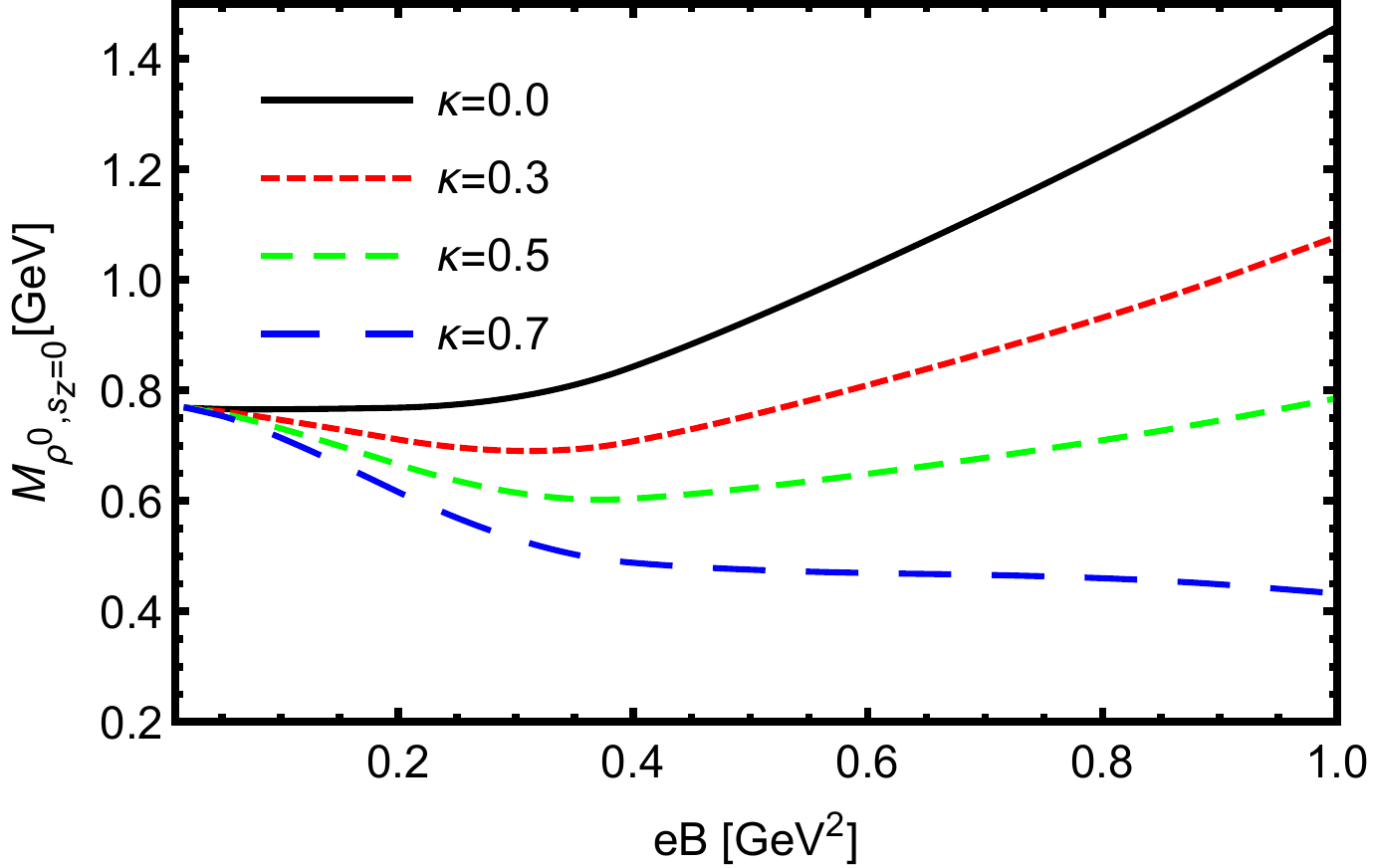}\label{subfig:neutralrhomassspin0}}
	\subfloat[]{\includegraphics[width=8cm]{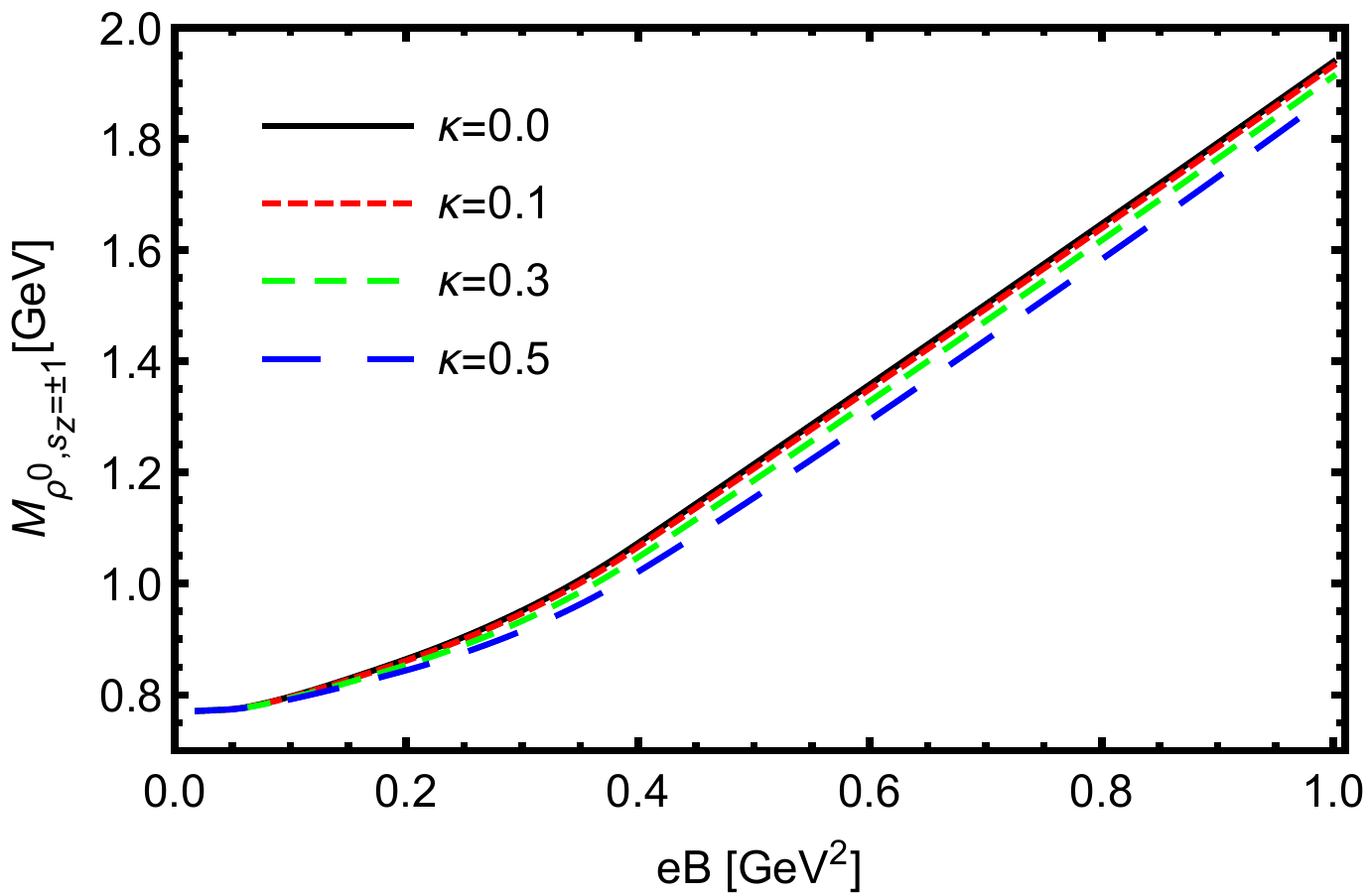}\label{subfig:neutralrhomassspin1}}
\caption{Mass of neutron rho with (a)$s_z=0$ and (b)$s_z=\pm 1$ as a function of magnetic field with different constant $\kappa$.}
\label{fig:nrhomass}
\end{figure}

Next, let's consider the effect of AMM on neutral rho mass with different spin component $s_z$, which is shown in Fig.(\ref{fig:nrhomass}). In the case of $\kappa=0$, all rho mass with three different $s_z$ increase with magnetic field, while the presence of AMM reduces neutral rho mass $M_{\rho^0}$ regardless its spin. However AMM can reduce mass of neutral rho with $s_z=0$ significantly while the mass of neutral rho with $s_z=\pm1$ is reduced slightly. Besides, when $\kappa \ge 0.7\text{GeV}^{-1}$, $M_{\rho^0}(s_z=0)$ decreases with magnetic field, opposite to the zero AMM case.

\begin{figure}[h!]
	\centering
	\subfloat[]{\includegraphics[width=8cm]{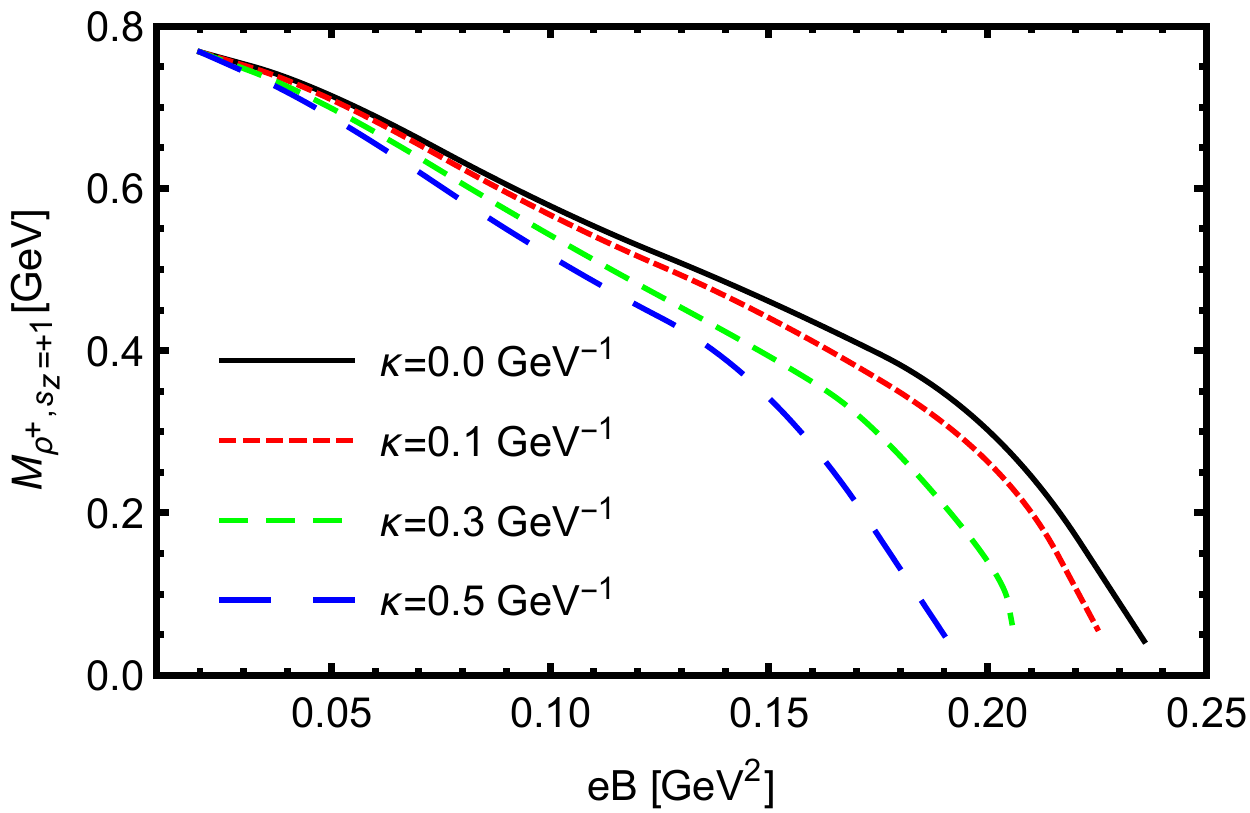}\label{subfig:chargedrhomassupspin}}
	\subfloat[]{\includegraphics[width=8cm]{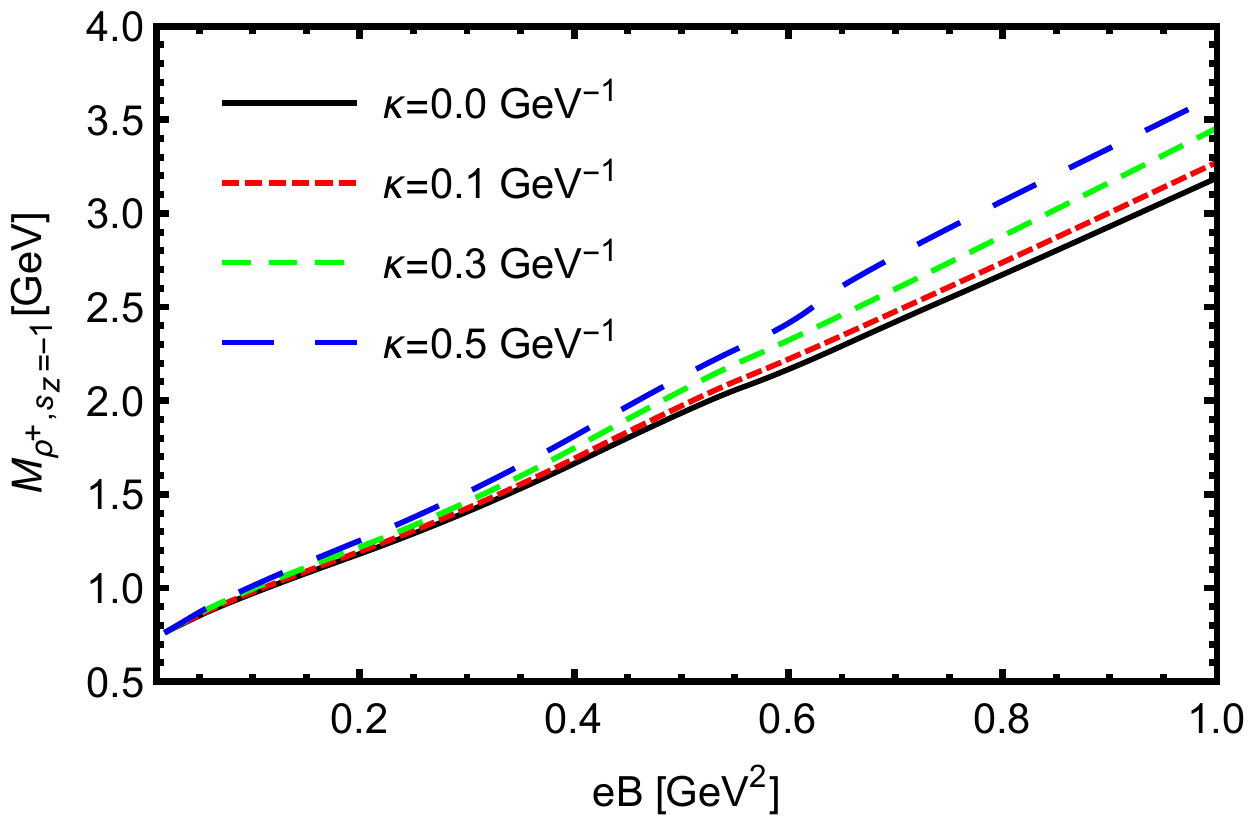}\label{subfig:chargedrhomassdownspin}}\\
	\subfloat[]{\includegraphics[width=8cm]{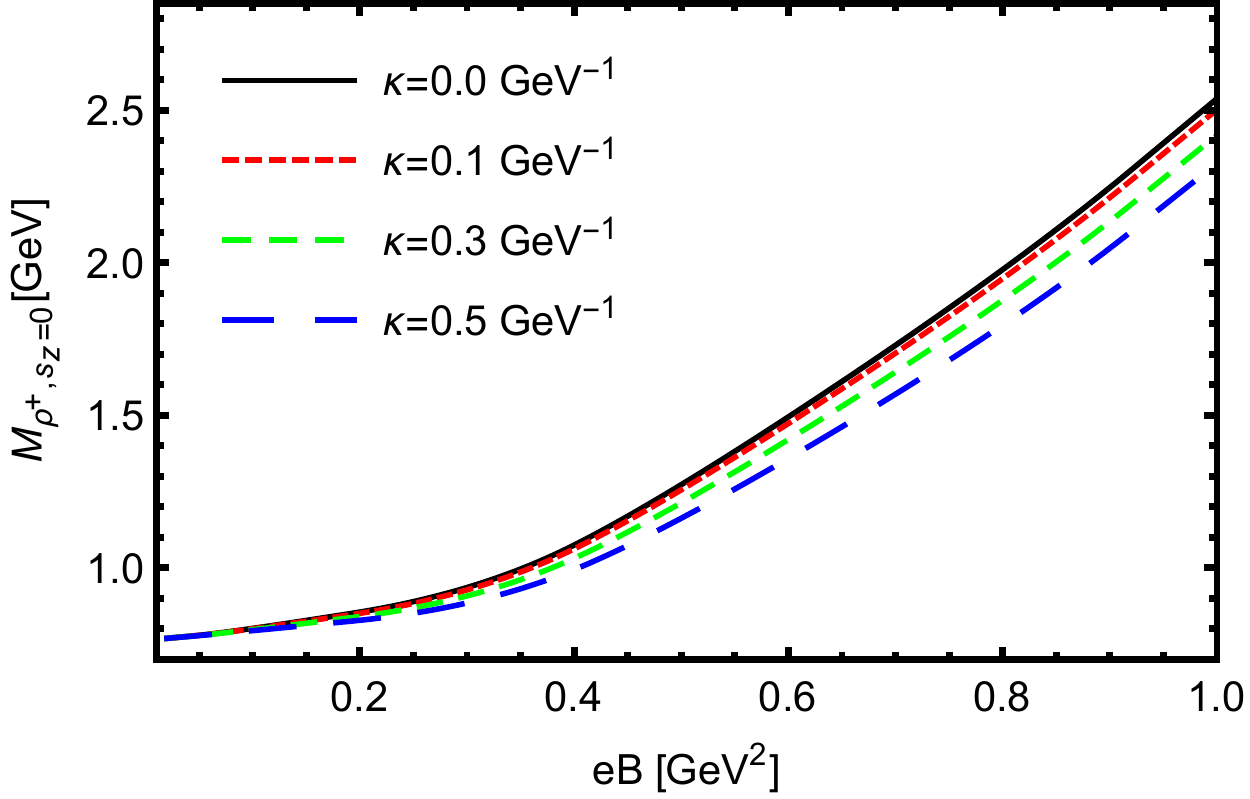}\label{subfig:chargedrhomassspin0}}
	\caption{Mass of charged rho with (a)$s_z=+1$, (b)$s_z=\pm 1$ and (c)$s_z=0$ as a function of magnetic field with  different constant $\kappa$.}
	\label{fig:crhomass}
\end{figure}

For charged rho we only consider it with positive charge, and the numerical results are shown in Fig.(\ref{fig:crhomass}). First of all, AMM reduces $M_{\rho^{+}}(s_z=+1)$ and $M_{\rho^{+}}(s_z=0)$ while increase $M_{\rho^{+}}(s_z=-1)$. For positive charged rho with $s_z=+1$, its mass $M_{\rho^{+}}(s_z=+1)$ decreases with magnetic in both zero and non-zero AMM cases, and drops to zero at a critical magnetic field $eB_c$, similar to neutral pion, $eB_c$ decreases as $\kappa$ increases, which indicates that the vacuum is more polarized when consider the AMM of quarks. For rho with $s_z=-1$ and $s_z=0$, the modifications of mass induced by AMM can be ignored compared to their mass.

\section{DISCUSSION AND Conclusion}
\label{sec-con}
In the present paper, we systematically investigate the effect of the anomalous magnetic moment(AMM) of quarks on the magnetized QCD matter in the two-flavor Nambu--Jona-Lasinio model, including the magnetic susceptibility, the inverse magnetic catalysis around the critical temperature and the neutral/charged pion and rho meson spectra under magnetic fields. 
The dynamical AMM of quarks is presented by a term $\bar{\psi} \kappa_f q_f \sigma^{\mu\nu}F_{\mu\nu}\psi$ in the Lagrangian, its coupling with magnetic field causes Zeeman splitting of the dispersion relation of quarks thus changes the magnetism properties and meson mass spectra under magnetic fields. It is found that though including the AMM of quarks cannot fully understand lattice results of the magnetized matter, it can partially explain lattice results. Our results can be summarized as following: 

1) The AMM of quarks reduces the dynamical quark mass which is consistent with the dispersion relation Eq.(\ref{equ:LLL}), a proper choice of $\kappa\sim \sigma $ can produce the inverse magnetic catalysis around the critical temperature $T_c$.  

2) The neutral pion mass is very sensitive to the AMM, it decreases with magnetic field quickly, and reaches zero at a critical magnetic field $eB_c$, and $eB_c$ decreases as $\kappa$ increases. On the other hand, the charged pion mass almost does not change with $\kappa$, it shows a nonlinear behavior, i.e., firstly linearly increases with the magnetic field and then saturates at strong magnetic field. The lattice calculation in \cite{Ding:2020jui} shows that the neutral pion mass decreases with the magnetic field and then saturate at about $0.6 M_{\pi^0}(eB=0)$ at large magnetic field $eB>1 {\rm GeV}^{2}$, and the charged pion mass firstly increases with magnetic field till $eB\sim 0.6 {\rm GeV}^{2}$ and then decreases with magnetic field. 

3) It's more complicated for $\rho$ because different spin component $s_z$ of rho behave differently. It is observed that AMM reduces the mass of neutral rho meson mass with different spin component $s_z$, and reduces the mass of $s_z=+1,0$ component charged rho meson mass but enhances the $s_z=-1$ component charged rho meson mass. Similar to pion, the neutral rho meson is more sensitive to the AMM of quarks than the charged rho.  AMM can reduce the mass of neutral neutral rho with $s_z=0$ and $s_z=\pm 1$, however the former one changes significantly while the latter one only shows a slight modification, which can be ignored compared to its mass. Besides, for $\kappa \ge 0.7 \text{GeV}^{-1}$, $M_{\rho^0}(s_z=0)$ decreases continuously with magnetic field, while it increases monotonously at zero AMM case.
For charged rho with spin component $s_z=+1$, its mass decreases with magnetic filed and drops to zero at a critical magnetic field $eB_c$, and the AMM reduces its mass as well as $eB_c$, which indicates that including the AMM of quarks makes the magnetized matter more easily to be polarized. Besides, AMM slightly reduces and enhances $M_{\rho^+}(s_z=0)$ $\rho^{+}(s_z=0)$ and $M_{\rho^+}(s_z=-1)$, respectively.

4) The magnetic susceptibility of the magnetized QCD matter cannot be understood by including the AMM of quarks. It is found that at low temperature the magnetic susceptibility can be either positive (paramagnetism) or negative(diamagnetism) with different AMM. 

In summary, taking into account of the AMM of quarks cannot fully explain lattice results on magnetized matter and meson spectra. More studies and other 
mechanism need to be investigated in the future.

\begin{acknowledgements}
	
This work is supported in part by the National Natural Science Foundation of China (NSFC)  Grant  Nos. 11735007, 11725523, and Chinese Academy of Sciences under Grant No. XDPB09, the start-up funding from University of Chinese Academy of Sciences(UCAS), and the Fundamental Research Funds for the Central Universities.

\end{acknowledgements}

\end{document}